\documentclass[letterpaper]{article} 
\usepackage[preprint]{aaai2027}

\usepackage[hyphens]{url}  
\usepackage{graphicx} 
\usepackage{natbib}  
\usepackage{caption} 
\usepackage{algorithm}
\usepackage{algorithmic}

\usepackage{newfloat}
\usepackage{listings}
\DeclareCaptionStyle{ruled}{labelfont=normalfont,labelsep=colon,strut=off} 
\floatstyle{ruled}
\newfloat{listing}{tb}{lst}{}
\floatname{listing}{Listing}

\usepackage{booktabs}

\usepackage{amsmath}
\usepackage{amssymb}
\usepackage{multirow}

\title{EEG2MOTION: Towards Open-Vocabulary Human Motion Synthesis from Non-invasive Brain Signals}
\author{
    Yulong Peng\textsuperscript{\rm 1,2}, Yijian Pan\textsuperscript{\rm 2}, Yuqi Yang\textsuperscript{\rm 2,3}, Nenggan Zheng\textsuperscript{\rm 2,3}, Weidong Chen\textsuperscript{\rm 2}, Xiaoling Hu\textsuperscript{\rm 4}, Shaomin Zhang\textsuperscript{\rm 2,5}\corresponding
}
\affiliations{
    \textsuperscript{\rm 1}Department of Biomedical Engineering, Zhejiang University\\
    \textsuperscript{\rm 2}Qiushi Academy for Advanced Studies, Zhejiang University\\
    \textsuperscript{\rm 3}College of Computer Science and Technology, Zhejiang University\\
    \textsuperscript{\rm 4}Department of Biomedical Engineering, The Hong Kong Polytechnic University\\
    \textsuperscript{\rm 5}Key Laboratory of Biomedical Engineering of Ministry of Education, Zhejiang University\\
    12515014@zju.edu.cn, 0925311@zju.edu.cn, yangyangyangyuqi@zju.edu.cn, zng@cs.zju.edu.cn, chenwd@zju.edu.cn, xiaoling.hu@polyu.edu.hk, shaomin@zju.edu.cn
}

\begin{document}

\maketitle

\begin{abstract}
Human motion is governed by a hierarchical motor system where the brain provides high-level intentions and lower-level structures coordinate detailed dynamics. Existing brain-computer interfaces (BCIs) typically oversimplify this into constrained classification or low-dimensional control, failing to capture the richness of natural movement. Bridging this gap to achieve open-vocabulary, full-body motion synthesis remains challenging due to the substantial cross-modal divergence between sparse neural signals and high-dimensional kinematics, as well as the lack of large-scale paired EEG-motion datasets. To address this, we introduce EEG2MOTION, the first EEG-motion-text dataset for human motion synthesis, comprising nearly 20,000 paired samples across thousands of motions. Using this dataset, we first demonstrate via multimodal contrastive learning that non-invasive EEG embeddings can be effectively aligned with text, video, and motion representations to decode high-level semantics. We then propose EEG-conditioned Masked Motion Model (EMMM), a generative framework that unites an EEG encoder with a motion decoder to synthesize continuous, full-body human motions directly from brain activity. Experimental results show that EMMM generates coherent and realistic motion sequences from non-invasive brain signals. To the best of our knowledge, this is the first work to generate diverse full-body human motions from non-invasive brain signals, opening a new direction toward generative and open-vocabulary motor BCIs. See our project page: \url{https://yulom.github.io/EEG2MOTIONdemopage/}.
\end{abstract}

\section{Introduction}
Human movement is governed by a hierarchical motor system that has evolved over millions of years. Many motor functions are distributed across lower-level neural structures, including subcortical regions, the brainstem, the spinal cord, and peripheral nervous systems, which establish sophisticated communication pathways with the cerebral cortex~\cite{frank1990coordination,rizzolatti2001cortical,shadmehr2008computational}. This biological architecture is particularly critical for the design of Brain-Computer Interfaces (BCIs)~\cite{wolpaw2007brain}. Because not all information required to reconstruct natural human motions is directly observable from cortical activity, motor BCIs require an intelligent ``copilot'' capable of receiving the multi-layered, sparse motor signals transmitted by the cortex and translating them into complete, natural human motions~\cite{lee2025brain}.

Most existing motor BCIs simplify movement decoding into highly constrained paradigms. Typical examples include classifying a small number of motor imagery (MI) categories or predicting two-dimensional and three-dimensional cursor trajectories~\cite{wang2012multi,wolpaw1991eeg}. While such paradigms have demonstrated practical utility, their limited motion space restricts deployment in complex real-world scenarios and provides only a partial understanding of the intricate brain-body interaction underlying human movement. In contrast to conventional constrained decoding tasks, we investigate, for the first time, the feasibility of open-vocabulary motion decoding from non-invasive electroencephalography (EEG) and further explore the synthesis of full-body human motions directly from brain activity.

Constructing an EEG data acquisition paradigm capable of synthesizing full-body motions presents substantial challenges. To mitigate these, we adopt a pragmatic paradigm centered on Action Observation (AO) rather than active motor execution or imagery~\cite{hardwick2018neural}. This design is motivated by several critical considerations. First, active motor execution inevitably introduces severe, hard-to-eliminate motion artifacts into EEG recordings. Furthermore, motor imagery is impractical for a motion space containing thousands of diverse actions and is often associated with large inter-subject variability~\cite{altaheri2023deep}. In contrast, visually evoked neural responses are relatively stable across participants, allowing efficient collection of large-scale datasets within a reasonable recording time~\cite{grootswagers2022human}. Neuroscience research has shown that action observation can elicit neural components that partially overlap with those involved in motor execution~\cite{hardwick2018neural}, and previous studies have demonstrated the feasibility of generating images and videos from EEG signals~\cite{bai2023dreamdiffusion,liu2024eeg2video}. Consequently, we designed an experimental paradigm where subjects wear EEG caps while observing videos of diverse human motions. We term this dataset EEG2MOTION. The motion samples are derived from the HumanML3D dataset~\cite{guo2022generating}, which contains text-to-motion data with each motion paired with one or more textual descriptions.

Leveraging this dataset, we conduct two core experiments. The first investigates multimodal contrastive learning~\cite{oord2018representation}, evaluating the feasibility of aligning EEG data with different modal representations. Textual descriptions, video frames, and motion posture sequences serve as the three target modalities. Experimental results demonstrate that EEG embeddings can be successfully aligned with representations from all three modalities and can generalize to previously unseen motions. 

Second, we propose a generative motion decoding framework based on masked motion model (MMM)~\cite{pinyoanuntapong2024mmm}. In this framework, EEG features extracted by an EEG encoder serve as conditioning signals for a motion generation model, while the motion decoder reconstructs human motion through masked token prediction. This formulation establishes a prototype of a generative motor decoding system: the brain provides sparse but informative motor cues, whereas the motion generation model acts as a ``motion copilot,'' learning a prior over human movements and translating neural intentions into complete and natural full-body motions.

In summary, the primary contributions of this work are three-fold:
\begin{itemize}
    \item \textbf{We introduce EEG2MOTION}, the first EEG dataset explicitly tailored for human motion synthesis, comprising nearly 20,000 paired EEG-motion-text samples and covering thousands of distinct human motions.
    \item \textbf{We benchmark multimodal contrastive learning on this dataset}, evaluating the alignment performance of EEG data against text, video, and motion features and achieving performance that significantly outperforms randomized baselines.
    \item \textbf{We propose EEG-conditioned Masked Motion Model (EMMM)}. By jointly optimizing an EEG encoder and a motion decoder, we achieve the first end-to-end synthesis of complete, natural human motions directly from non-invasive EEG signals.
\end{itemize}

\section{Related Work}

\subsection{Text-to-Motion Generation}
Benefiting from the emergence of large-scale annotated benchmarks such as KIT ~\cite{plappert2016kit} and HumanML3D ~\cite{guo2022generating}, the field of cross-modal motion synthesis has witnessed rapid advancement. Contemporary methodologies primarily bifurcate into continuous Diffusion-based models ~\cite{tevet2022human,karunratanakul2023guided,zhang2024motiondiffuse} and discrete Autoregressive or Masked Transformers ~\cite{pinyoanuntapong2024mmm,pinyoanuntapong2024bamm, jiang2023motiongpt}. Beyond pure text-conditioned generation, substantial efforts have focused on fine-grained motion editing and the integration of complex conditioning factors, such as precise joint trajectory guidance ~\cite{pinyoanuntapong2025maskcontrol} and dynamic scene interactions ~\cite{tevet2025closd}. These advancements not only propel human motion synthesis toward real-world applications but also extract rich, hierarchical behavioral representations from vast datasets. Crucially, they shed light on the structural grammar of body language: identifying which critical semantic components must be actively commanded by the brain, and which peripheral details can be naturally complemented by learned motor priors. This computational insight creates immense, yet largely untapped, potential for integration with brain-computer interfaces.

\subsection{Generative Brain Decoding}
In recent years, neural decoding has shifted from constrained classification and regression toward Generative Brain Decoding~\cite{shukla2025survey}. In this paradigm, neural features are aligned with target modality representations to form a shared latent space, while pretrained generative foundation models provide modality priors that enable the synthesis of natural content from sparse neural cues.

Recent breakthroughs span both invasive and non-invasive BCIs across various domains, creating generative BCIs counterparts whenever a powerful foundation model emerges, such as in image synthesis~\cite{bai2023dreamdiffusion,kavasidis2017brain2image}, video reconstruction~\cite{liu2024eeg2video}, speech synthesis~\cite{littlejohn2025streaming,chen2024neural,anumanchipalli2019speech}, or open-vocabulary text translation~\cite{liu2024eeg2text,wang2022open}. 
Crucially, within the motor decoding domain, a parallel shift toward unconstrained, whole-body generation has recently begun; He~et~al.~\cite{he2026neural} successfully decoded unconstrained whole-body kinematics in freely moving monkeys by conditioning an autoregressive generative model on intracranial cortical signals. 
However, the integration of generative human motion models with non-invasive EEG decoding remains entirely unexplored. Our work bridges this critical gap, extending open-vocabulary generative neural decoding into the realm of complex human kinematics.

\section{EEG2MOTION Dataset}

\begin{figure*}[t]
\centering
\includegraphics[width=0.9\textwidth]{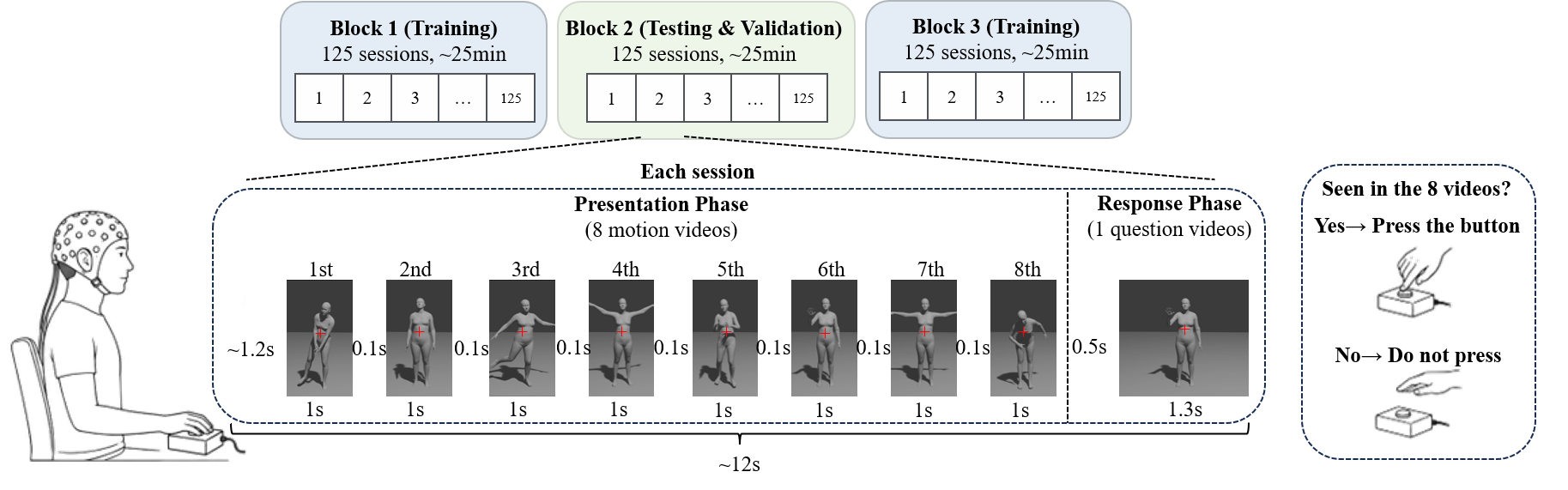} 
\caption{Illustration of the EEG2MOTION data collection paradigm. Participants wearing an EEG cap observed human motion videos rendered from HumanML3D motion sequences while EEG signals were continuously recorded. The experiment consisted of three blocks, each containing 125 sessions. The second block was reserved for validation and testing, whereas the remaining blocks were used for training. Each session contained 8 motion clips of one second each. Participants were instructed to maintain visual attention and indicate whether they remained attentive using a button press after each session.}
\label{fig_paradigm}
\end{figure*}

\begin{figure}[t]
\centering
\includegraphics[width=0.9\columnwidth]{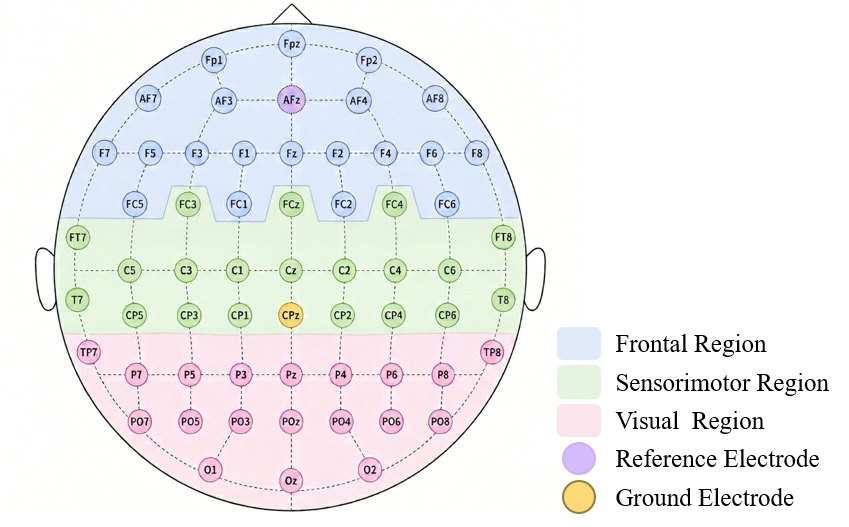} 
\caption{EEG electrode montage and regional partition. The 59-channel EEG setup followed the international 10–10 system. Electrodes were divided into three approximately balanced groups according to their spatial locations: frontal, sensorimotor, and posterior visual regions, corresponding to high-level action understanding, motor representation, and visual motion perception processes, respectively. AFz and CPz were used as the reference and ground electrodes.}
\label{fig_electrode}
\end{figure}

\subsection{Data Acquisition Paradigm}
Our experimental paradigm is shown in Fig.~\ref{fig_paradigm}. Continuous scalp EEG was collected from 9 subjects using a 64-channel recording system at 1000~Hz, providing 59 channels after excluding channels like EOG (see Fig.~\ref{fig_electrode}). Each subject completed 3 recording blocks (one subject had an additional training block); block 2 was used for validation and testing, and the rest for training. 

Subjects were seated comfortably and instructed to fixate on a central red cross to minimize eye movements. Visual stimuli consisted of motion data selected from the HumanML3D dataset~\cite{guo2022generating}, which were then rendered using Blender. Although each motion sequence is paired with textual descriptions in the dataset, only the videos were presented during the experiment. Motions with original durations of 4--6 seconds were selected and standardized to 5 seconds via truncation or last-frame delay, and played at $5\times$ speed, yielding 1 second per motion clip. The training, validation, and test samples were drawn exclusively from corresponding HumanML3D splits to ensure test and validation motions were unseen during training.

Each block comprised 125 sessions. Each session lasted approximately 12 seconds, containing 8 sequential motion clips, followed by a 1.8-second response phase where subjects identified a query motion. This task was used solely to maintain visual attention, and data from the response phase was excluded from analysis. Consequently, each block yielded 1,000 paired EEG-motion-text samples (block 2 provided 250 validation and 750 test samples per subject).

\subsection{Data Preprocessing}
\subsubsection{EEG Preprocessing.} Each 1-second EEG epoch was band-pass filtered from 0.5 to 45~Hz and downsampled to 250~Hz. Channel-wise z-score normalization was applied using parameters from each subject's training set. The 59 electrodes were partitioned into three functionally balanced regions (see Fig.~\ref{fig_electrode}): frontal, sensorimotor, and visual regions. The final EEG tensor shape is $C \times 250$ ($C = 59$).

\subsubsection{Motion Preprocessing.} Human motion followed the standard 263-dimensional representation in HumanML3D \cite{guo2022generating}, encapsulating positions, velocities of 23 joints, and foot-ground contacts. Played back at $5\times$ speed from the original 20 FPS, each motion sequence spans a maximum of 100 frames.

\subsubsection{Subject Screening and Dataset Compilation.} A preliminary experiment was conducted to assess each subject's EEG-motion contrastive alignment (see Supplementary Material). We excluded three subjects whose EEG representations failed to achieve significant contrastive alignment with motion features compared to randomized noise inputs. Data from the remaining subjects were merged into a unified dataset encompassing 13,000 training, 1,500 validation, and 4,500 test samples.

\section{Method}

\subsection{Multimodal Contrastive Learning}
\begin{figure*}[t]
\centering
\includegraphics[width=0.9\textwidth]{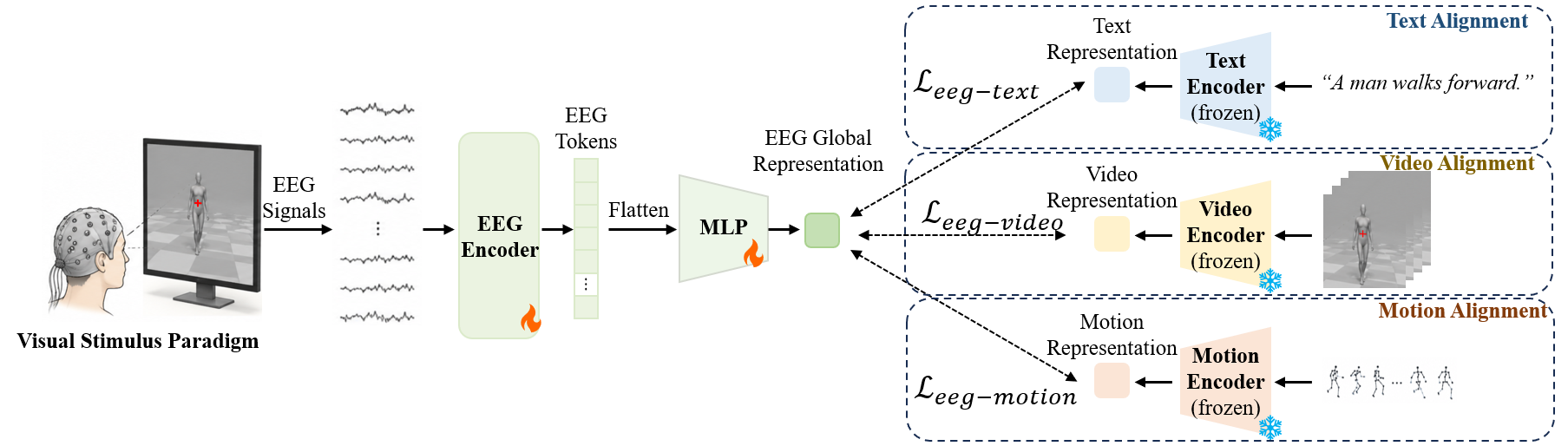} 
\caption{An overview of the proposed cross-modal contrastive learning framework. Continuous EEG signals are encoded into neural tokens via EEGConformer and mapped to the joint latent space using an MLP head. Frozen pre-trained target encoders (CLIP, VideoMAE, and Motion Encoder) extract target features. A symmetric InfoNCE loss is employed to align EEG features and target features}
\label{fig_contrast}
\end{figure*}

To investigate whether EEG signals can be aligned with diverse modalities in an open-vocabulary scenario, we propose a cross-modal contrastive learning framework (Fig.~\ref{fig_contrast}).

We sequentially adopt text, video, and human motion as the target modality. Frozen pre-trained feature extractors---specifically the CLIP text encoder~\cite{radford2021learning}, VideoMAE~\cite{tong2022videomae}, and a motion encoder pre-trained via contrastive learning with text~\cite{guo2022generating}---are leveraged to extract target features $\mathbf{Z}_{\text{target}} \in \mathbb{R}^{B \times D}$ ($\text{target} \in \{\text{text}, \text{video}, \text{motion}\}$), where $B$ and $D$ denote the batch size and feature dimension, respectively. The size of $D$ depends on the output feature dimension of the corresponding encoder. For VideoMAE, the features are obtained by averaging the output spatiotemporal tokens.

For the EEG modality, we employ the EEGConformer as the feature extractor~\cite{song2022eeg}, where the batch normalization layers are replaced with subject-specific batch normalization to alleviate inter-subject variability. The raw EEG input $\mathbf{X} \in \mathbb{R}^{B \times C \times T}$ ($C$: channels, $T$: time points) is transformed via a convolutional embedding layer and processed by Transformer blocks to yield the EEG tokens $\mathbf{E}_{\text{eeg}} \in \mathbb{R}^{B \times K \times N}$ ($K$: token number, $N$: token dimension). 
These tokens are then flattened and forwarded through an MLP projection head to produce the final EEG global features $\mathbf{Z}_{\text{eeg}} = \text{MLP}(\text{Flatten}(\mathbf{E}_{\text{eeg}})) \in \mathbb{R}^{B \times D}$.

To achieve cross-modal alignment, we apply a symmetric InfoNCE loss on the $L_2$-normalized features $\hat{\mathbf{Z}}_{\text{eeg}}$ and $\hat{\mathbf{Z}}_{\text{target}}$~\cite{radford2021learning}:

\begin{equation}
\begin{aligned}
\mathcal{L}_{\text{align}} = -\frac{1}{2B} \sum_{i=1}^{B} & \left( \log \frac{\exp(\mathbf{S}_{i,i}/\tau)}{\sum_{j=1}^{B} \exp(\mathbf{S}_{i,j}/\tau)} \right. \\
& \left. + \log \frac{\exp(\mathbf{S}_{i,i}/\tau)}{\sum_{j=1}^{B} \exp(\mathbf{S}_{j,i}/\tau)} \right)
\end{aligned}
\end{equation}

where $\mathbf{S} = \hat{\mathbf{Z}}_{\text{eeg}} \hat{\mathbf{Z}}_{\text{target}}^{\top} \in \mathbb{R}^{B \times B}$ represents the cosine similarity matrix, and $\tau$ is a temperature parameter (set to $0.07$ empirically) to scale the logits.

\subsection{EEG-conditioned Masked Motion Model}
\begin{figure*}[t]
\centering
\includegraphics[width=0.9\textwidth]{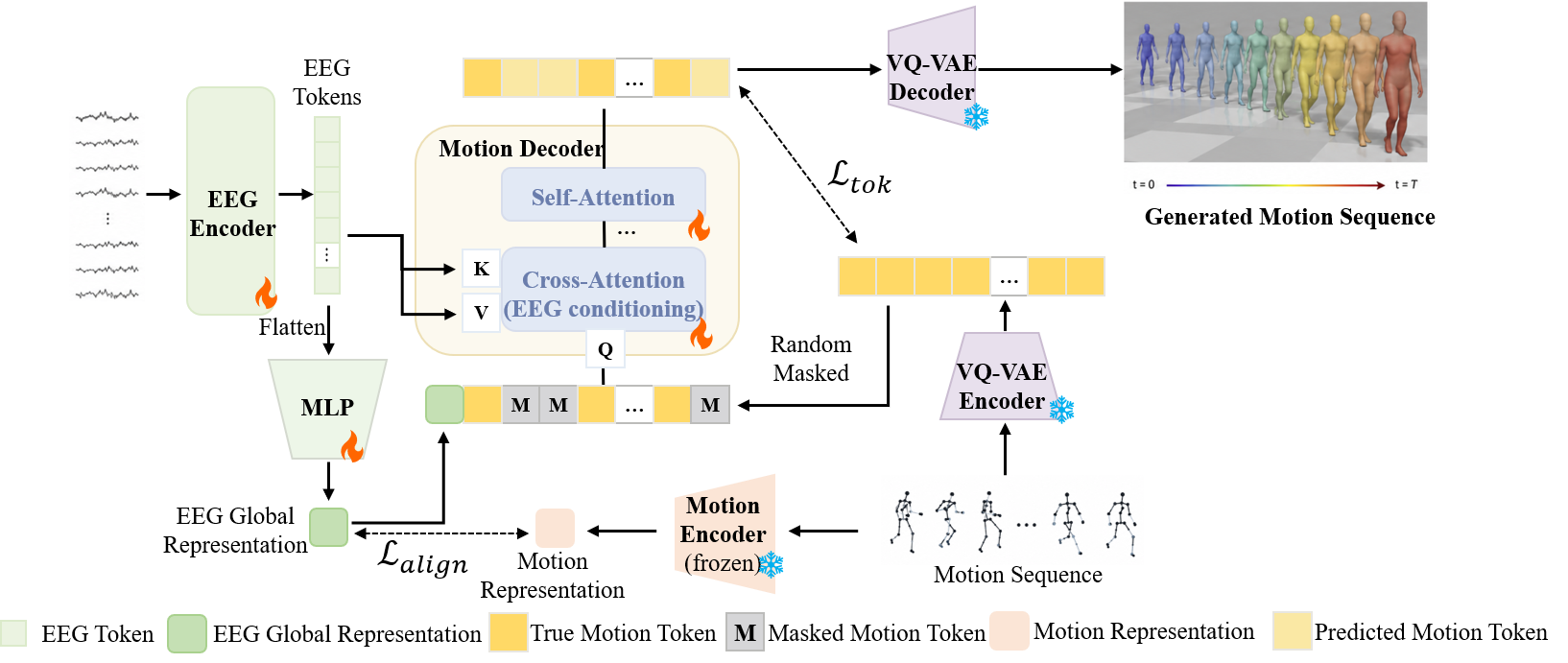} 
\caption{Overview of the EEG-conditioned Masked Motion Model framework. The architecture consists of an EEG encoder, a pre-trained VQ-VAE, and a Transformer-based motion decoder. Inside the decoder, global EEG features are prepended to motion tokens as a prompt prefix, while fine-grained EEG tokens provide neural guidance via cross-attention layers. During training, the alignment loss $\mathcal{L}_{\text{align}}$ is computed between the global EEG features and the target motion features. Concurrently, the ground-truth motion tokens are randomly masked and fed alongside the EEG inputs into the motion decoder, which is optimized via $\mathcal{L}_{\text{tok}}$ to predict the codebook indices corresponding to the masked positions.}
\label{fig_generation}
\end{figure*}

To synthesize realistic human motions directly from neural representations, we propose an EEG-conditioned Masked Motion Model framework (Fig.~\ref{fig_generation}). We integrate the EEG encoder with the motion decoder from MMM~\cite{pinyoanuntapong2024mmm}, a text-to-motion generation model, and train them together to enable EEG-conditioned human motion synthesis.

\subsubsection{Model Architecture.}
The architecture comprises an EEG encoder, a pre-trained VQ-VAE for motion tokenization, and a Transformer-based motion decoder. The EEG encoder adopts the identical configuration as the contrastive learning section but is trained from scratch. Given the raw EEG input $\mathbf{X}$, it extracts the fine-grained EEG tokens $\mathbf{E}_{\text{eeg}} \in \mathbb{R}^{B \times K \times N}$ and the global EEG representation $\mathbf{Z}_{\text{eeg}} \in \mathbb{R}^{B \times D}$. 

Following~\cite{pinyoanuntapong2024mmm}, the pre-trained VQ-VAE encoder compresses every 4 frames of continuous human motion into a single discrete token, yielding a target sequence $\mathbf{Y} = \{y_i\}_{i=1}^{L}$ of length $L$, while its decoder reconstructs continuous motions from these tokens. To effectively inject neural guidance, the global EEG feature $\mathbf{Z}_{\text{eeg}}$ is first prepended to the motion token sequence as a prompt prefix. This combined sequence is then fed into the Transformer-based motion decoder. 

In the initial layers of the motion decoder, a cross-attention mechanism aligns motion queries with neural contexts, formulated as $\text{Attention}(\mathbf{Q}_m, \mathbf{K}_e, \mathbf{V}_e) = \text{softmax}(\mathbf{Q}_m \mathbf{K}_e^{\top} / \sqrt{D})\mathbf{V}_e$, where $\mathbf{Q}_m$ is derived from the motion tokens, and $\mathbf{K}_e, \mathbf{V}_e$ are projected from $\mathbf{E}_{\text{eeg}}$. The subsequent layers utilize self-attention to model temporal motion dynamics. Learnable special tokens, including $\text{[MASK]}$, $\text{[PAD]}$, and $\text{[END]}$, are employed for masking, padding, and indicating the termination of the motion sequence.

\subsubsection{Training Strategy and Objectives.}
To regulate the information bottleneck between the EEG encoder and the motion decoder, the symmetric alignment loss $\mathcal{L}_{\text{align}}$ from the contrastive section remains active. 

During training, a specific percentage of tokens in $\mathbf{Y}$ are randomly replaced with the $\text{[MASK]}$ token, yielding the partially corrupted sequence $\mathbf{Y}_{\bar{\mathcal{M}}}$. The motion decoder is optimized to predict the codebook indices of the masked tokens conditioned on the unmasked context and the EEG signals. The token prediction loss $\mathcal{L}_{\text{tok}}$ is defined as:

\begin{equation}
\mathcal{L}_{\text{tok}} = -\mathbb{E} \left[ \sum_{i \in \mathcal{M}} \log p(y_i \mid \mathbf{Y}_{\bar{\mathcal{M}}}, \mathbf{E}_{\text{eeg}}, \mathbf{Z}_{\text{eeg}}) \right]
\end{equation}

where $\mathcal{M}$ denotes the set of masked token indices. 

To preserve the prior knowledge of the generative model, the weights of the motion decoder are frozen, while LoRA is deployed to fine-tune the $\mathbf{Q}, \mathbf{K}, \mathbf{V}$ projection matrices in the cross-attention layers and the $\mathbf{Q}, \mathbf{V}$ matrices in the self-attention layers~\cite{hu2022lora}. The EEG encoder is fully trainable.

\subsubsection{Inference Stage.}
During inference, the motion sequence length is fixed to $L = 25$ tokens (corresponding to 1 second of motion video) and entirely initialized with $\text{[MASK]}$ tokens. Conditioned on the EEG input, a progressive iterative decoding strategy identical to MMM is executed. In each iteration, the model predicts the probability distribution over all masked positions, retains a subset of tokens with the highest confidence, and re-masks the remaining uncertain positions at a specific ratio until the complete sequence is recovered. The finalized token sequence is forwarded to the VQ-VAE decoder to synthesize the final motion sequence.

\section{Experiments}
\subsubsection{Implementation Details.}

We evaluate our framework on the EEG2MOTION dataset. The EEG encoder configuration follows EEGConformer~\cite{song2022eeg}, and the motion decoder inherits hyperparameters from MMM~\cite{pinyoanuntapong2024mmm} (with the text encoder removed); detailed specifications are provided in the Supplementary Material. 

For motion synthesis, we compare against two additional strategies. The first is training a motion decoder from scratch, which uses the same architecture but with fewer layers and reduced hidden dimensions to prevent overfitting given data and hardware constraints. The second is a knowledge distillation strategy~\cite{hinton2015distilling}, where the original text-to-motion model MMM serves as the teacher. Specifically, the same masked motion token sequences are fed into both our EEG-to-motion model and MMM, and a KL divergence distillation loss is computed between the output distributions at the corresponding token positions.

All models are trained for $10,000$ iterations on two NVIDIA RTX 3090 GPUs using the AdamW optimizer. The learning rate starts at $2 \times 10^{-4}$ and decays to $1 \times 10^{-4}$ after $8,000$ iterations. The optimal checkpoint is selected via validation performance. The default batch size is $512$, which is adjusted to $256$ during LoRA fine-tuning of the MMM decoder due to memory limits. For motion synthesis, loss coefficients for $\mathcal{L}_{\text{align}}$ and $\mathcal{L}_{\text{tok}}$ are empirically set to $1$.

\subsubsection{Evaluation Metrics.}

Operating in open-vocabulary scenarios, we adopt \textbf{R-Precision} as our primary metric. During testing, for each trial, $32$ candidate targets (one matched ground-truth and $31$ randomly selected distractors) are sampled from the test set. For EEG-to-text/video/motion retrieval, we compute cosine similarity between the EEG query and each candidate target, and rank the candidates accordingly. For motion synthesis, we compute the feature distance between the synthesized motion and $32$ candidate motions, and rank them in ascending order. The Top-1, Top-2, and Top-3 accuracies (whether the ground-truth target ranks within the top $k$) are reported, averaged over multiple runs.

For motion synthesis, we further incorporate three established generative metrics from text-to-motion research~\cite{guo2022generating}. \textbf{Fr{\'e}chet Inception Distance} (FID) measures the distribution distance between the feature of synthesized motions and real motions. \textbf{Multimodal Distance (MM-Dist)} measures the distance in feature space between synthesized motions and their paired ground-truth motions. \textbf{Diversity} calculates the average distance in feature space across $300$ randomly sampled pairs of synthesized motions.

\begin{table*}[t]
\centering
\begin{tabular}{llccc}
\toprule
\textbf{Evaluation Setting} &
\textbf{Channels} &
\textbf{Top1} $\uparrow$&
\textbf{Top2} $\uparrow$&
\textbf{Top3} $\uparrow$\\
\midrule

shuffled baseline
& all
& 0.033$^{\pm.004}$
& 0.064$^{\pm.004}$
& 0.097$^{\pm.005}$ \\

noise baseline
& all
& 0.030$^{\pm.003}$
& 0.062$^{\pm.004}$
& 0.094$^{\pm.006}$ \\
\midrule

eeg-text
& all
& 0.035$^{\pm.002}$
& 0.074$^{\pm.004}$
& 0.106$^{\pm.005}$ \\
\midrule

\multirow{2}{*}{eeg-video}
& visual only
& \textbf{0.047}$^{\pm.004}$
& \textbf{0.090}$^{\pm.005}$
& \textbf{0.132}$^{\pm.005}$ \\

&
all
& 0.037$^{\pm.004}$
& 0.073$^{\pm.005}$
& 0.109$^{\pm.005}$ \\
\midrule

\multirow{4}{*}{eeg-motion}

& frontal only
& 0.034$^{\pm.002}$
& 0.068$^{\pm.004}$
& 0.103$^{\pm.004}$ \\

& sensorimotor only
& 0.038$^{\pm.003}$
& 0.072$^{\pm.004}$
& 0.102$^{\pm.006}$ \\

& visual only
& \underline{0.044}$^{\pm.003}$
& 0.086$^{\pm.005}$
& \underline{0.126}$^{\pm.007}$ \\

& all
& 0.041$^{\pm.003}$
& \underline{0.087}$^{\pm.006}$
& 0.125$^{\pm.006}$ \\

\bottomrule
\end{tabular}

\caption{Retrieval accuracy of multimodal contrastive learning. Values are reported as mean with 95\% confidence interval.}

\label{tab:contrastive}

\end{table*}

\begin{table*}[t]
\centering
\begin{tabular}{llllcccc}
\toprule
\textbf{Evaluation Setting} &
$\boldsymbol{\mathcal{L}}_{\text{align}}$ &
\textbf{Channels} &
\textbf{Method} &
\textbf{Top3} $\uparrow$ &
\textbf{MM-Dist} $\downarrow$ &
\textbf{FID} $\downarrow$ &
\textbf{Diversity} $\uparrow$ \\
\midrule

shuffled baseline
& eeg-motion
& all
& finetuning
& 0.091$^{\pm.005}$
& 9.815$^{\pm.079}$
& -
& - \\

noise baseline
& eeg-motion
& all
& finetuning
& 0.092$^{\pm.006}$
& 9.638$^{\pm.074}$
& 8.360$^{\pm.601}$
& 8.668$^{\pm.112}$ \\
\midrule

\multirow{8}{*}{motion retrieval}

& none
& all
& from scratch
& 0.106$^{\pm.004}$
& 9.392$^{\pm.049}$
& \textbf{2.324}$^{\pm.143}$
& 9.199$^{\pm.085}$ \\
\cmidrule{2-8}

& \multirow{2}{*}{eeg-video}
& \multirow{2}{*}{all}
& from scratch
& \textbf{0.123}$^{\pm.007}$
& \textbf{9.119}$^{\pm.071}$
& 7.130$^{\pm.324}$
& 8.926$^{\pm.147}$ \\

&
&
& finetuning
& 0.110$^{\pm.005}$
& 9.447$^{\pm.053}$
& 5.721$^{\pm.279}$
& 9.298$^{\pm.125}$ \\
\cmidrule{2-8}

& \multirow{5}{*}{eeg-motion}
& \multirow{2}{*}{visual only}
& from scratch
& 0.119$^{\pm.005}$
& 9.299$^{\pm.046}$
& 10.641$^{\pm.523}$
& 8.724$^{\pm.136}$ \\

&
&
& finetuning
& 0.120$^{\pm.006}$
& 9.324$^{\pm.063}$
& 6.466$^{\pm.368}$
& 9.252$^{\pm.086}$ \\
\cmidrule{3-8}

&
& \multirow{3}{*}{all}
& from scratch
& \underline{0.121}$^{\pm.005}$
& \underline{9.258}$^{\pm.053}$
& 2.825$^{\pm.264}$
& \textbf{9.471}$^{\pm.111}$ \\

&
&
& distillation
& 0.112$^{\pm.005}$
& 9.558$^{\pm.048}$
& 17.354$^{\pm.585}$
& 8.909$^{\pm.126}$ \\

&
&
& finetuning
& 0.119$^{\pm.006}$
& 9.275$^{\pm.056}$
& \underline{2.743}$^{\pm.221}$
& \underline{9.411}$^{\pm.097}$ \\

\bottomrule
\end{tabular}

\caption{Performance of motion synthesis under different evaluation settings. Values are reported as mean with 95\% confidence interval.}

\label{tab:sythesis_motion}

\end{table*}

\begin{table}[t]
\centering
\small
\begin{tabular}{lccc}
\toprule
\textbf{Evaluation Setting} &
\textbf{Encoder Only} &
\textbf{+ Decoder} &
\textbf{Gain} \\
\midrule

eeg-video
&
0.109$^{\pm0.005}$
&
0.144$^{\pm0.005}$
&
+ 32.1\% \\

eeg-motion
&
0.125$^{\pm0.006}$
&
0.143$^{\pm0.005}$
&
+ 14.4\% \\

\bottomrule
\end{tabular}

\caption{
Retrieval Top3 accuracy of EEG global representation and target modality features. Adding a motion decoder improves the alignment.
}

\label{tab:decoder_gain}

\end{table}

\begin{figure*}[t]
\centering
\includegraphics[width=0.9\textwidth]{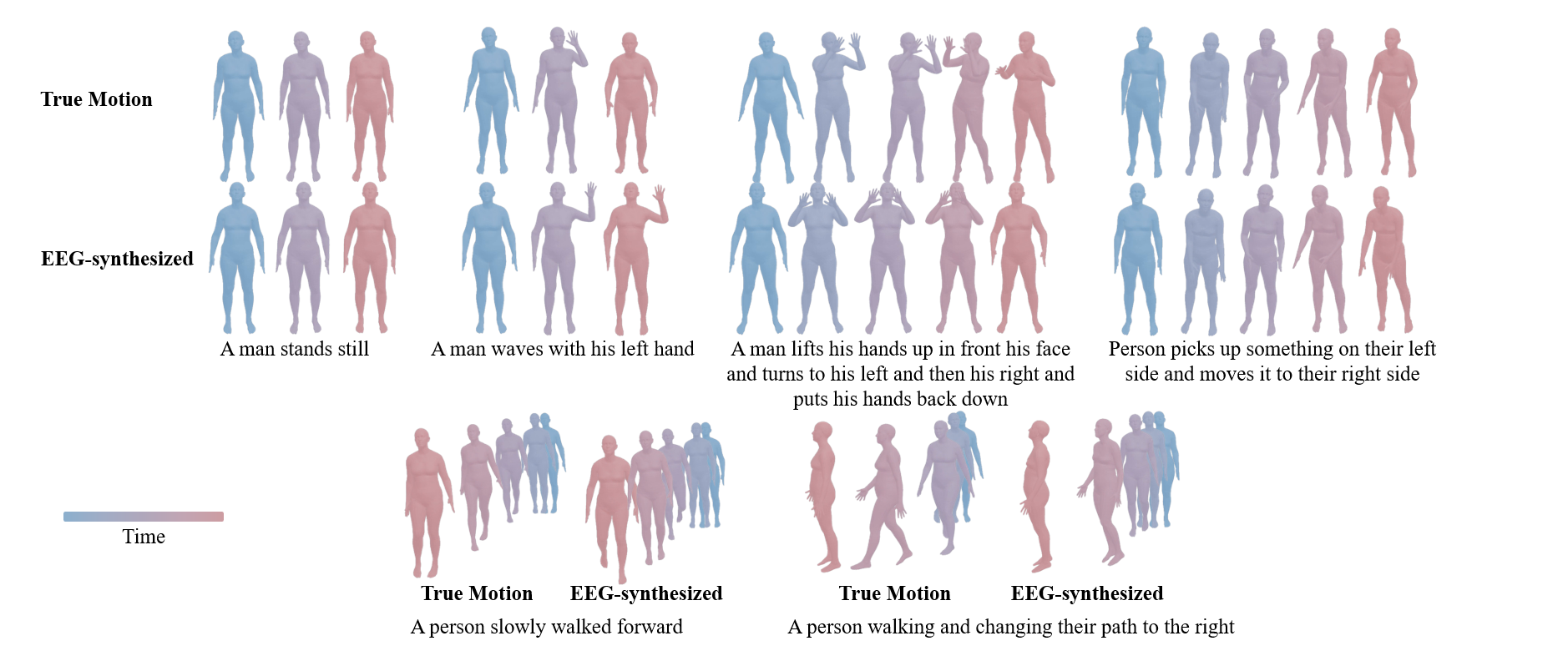} 
\caption{Examples of human motions synthesized from EEG.}
\label{fig_motion}
\end{figure*}

To address concerns that a powerful decoder might synthesize plausible motions independently, rendering EEG conditioning redundant, we evaluate two randomized control baselines: (1) a \textbf{Gaussian Noise Baseline} where input EEG is replaced with standard Gaussian noise, and (2) a \textbf{Shuffled Baseline} where EEG inputs are randomly paired with targets.

\section{Results}
\subsection{Multimodal Contrastive Learning Results}
The cross-modal retrieval results are summarized in Table~\ref{tab:contrastive}. All three target modalities achieve retrieval accuracies that significantly outperform the random chance levels (detailed statistical tests are in the Supplementary Material). To investigate the spatial contributions of neural dynamics, we conduct a channel ablation study on the video and motion retrieval tasks by comparing different channel configurations.

The empirical results reveal that channels covering the visual cortex play a mayor role in cross-modal alignment, occasionally surpassing the full-channel configuration. This suggests that under our current paradigm, visual cortical activations encode substantial task-related semantics driven by action observation mechanisms. Conversely, sensorimotor channels yield relatively limited retrieval capabilities, potentially because the rapid, passive video-viewing task inherently prioritizes visuospatial processing over active motor engagement. Nevertheless, these findings demonstrate that the joint latent space successfully bridges neural signals with open-vocabulary motion semantics, providing a solid foundation for subsequent motion synthesis.

\subsection{Motion Synthesis Results}
The evaluation of the EMMM framework is summarized in Table~\ref{tab:sythesis_motion}. 
To facilitate quantitative evaluation, the synthesized and ground-truth motions are projected into a shared latent space using a pre-trained motion encoder, same with text-to-motion generation research~\cite{guo2022generating}. 
We comprehensively ablate the effects of the alignment objective $\mathcal{L}_{\text{align}}$, channel combinations, and training methods.

The results reveal that removing $\mathcal{L}_{\text{align}}$ leads to a severe degradation in retrieval performance, despite achieving the lowest FID score. 
This phenomenon highlights a trade-off between synthesis realism and conditional fidelity; without $\mathcal{L}_{\text{align}}$, the motion decoder can still synthesize plausible human movements, but it loses conditional dependency on EEG, validating that $\mathcal{L}_{\text{align}}$ acts as a crucial information bottleneck to constrain semantic extraction.

Computing $\mathcal{L}_{\text{align}}$ via video features, motion features, or utilizing visual-only channels yields comparable retrieval accuracy. However, combining all channels with motion feature alignment produces a substantially lower FID and higher diversity. This indicates that while visuospatial information provides adequate cross-modal matching cues, achieving realistic and diverse motion synthesis ultimately requires the integration of sensorimotor representations. 

Regarding training strategies, LoRA fine-tuning and training from scratch achieve comparable performance, whereas a naive knowledge distillation baseline underperforms, suggesting that cross-modal distillation from established text-to-motion priors demands specialized adaptation frameworks for neural signals.

We also evaluate the retrieval performance between the global EEG representations and the target modality (Table~\ref{tab:decoder_gain}). 
Intriguingly, joint optimization with the downstream motion decoder significantly enhances the retrieval accuracy of the global EEG representation, outperforming all results from the isolated contrastive learning phase. This indicates a cross-level reciprocal optimization: fine-grained motion reconstruction constraints provide complementary low-level details that enrich high-level neural semantic capture, while motion reconstruction concurrently benefits from tighter cross-modal alignment.

Finally, Fig.~\ref{fig_motion} qualitative visualizes successful retrieval exemplars showcasing the ground-truth and EEG-synthesized motions. Our framework successfully generates realistic, diverse, and semantically aligned human motions guided by non-invasive neural signals. Additional qualitative results are detailed in the Supplementary Material.

\section{Discussion and Limitations}
While EMMM achieves the first full-body human motion synthesis from non-invasive EEG, several limitations remain. 
First, our Action Observation paradigm implies that the retrieval performance is largely driven by visual cortical activations rather than the sensorimotor regions. Whether the learned representations can transfer to active control paradigms remains to be validated. 
Second, although our results significantly outperform randomized baselines, a substantial performance gap exists compared to text-to-motion generation, primarily due to the inherent low signal-to-noise ratio of EEG, which limits its practical applications. Third, our framework prioritizes high-level semantic retrieval over absolute joint trajectories, leading to spatial precision deviations. Integrating a precise joint trajectory controller into the generative backbone is a promising. Finally, our model lacks detailed modeling of hand keypoints. Incorporating hand kinematics is essential for building more practical BCI systems.

Nevertheless, this work represents an initial step toward generative motor BCIs. While we utilize non-invasive EEG as an accessible starting point, the proposed open-vocabulary generative decoding framework may also provide insights for the development of invasive motor BCIs, which may offer greater potential for practical applications. We believe that generative motion decoding based on invasive neural recordings will emerge as a key research direction and application paradigm for next-generation motor BCIs.

\section{Conclusion}
In this work, we introduced EMMM, a generative framework that successfully synthesizes realistic, full-body human motions from non-invasive EEG signals. 
Supported by our curated EEG2MOTION dataset, we demonstrated that integrating cross-modal alignment with downstream motion modeling effectively balances generative realism and conditional fidelity. 
By bridging neural signals with high-dimensional kinematics, this framework establishes a competitive baseline and opens a promising frontier for future generative and open-vocabulary motor BCIs.

\bibliography{aaai2027}


\end{document}